\documentclass[twocolumn,aps,prb,showpacs,amsmath]{revtex4}

\usepackage{graphicx}

\begin{document}

\title{
On Feynman's calculation of the Fr\"ohlich polaron mass
}

\author{Pavel Kornilovitch}
\affiliation{
2876 N.W. Audene Drive, Corvallis, Oregon, 97330, USA
}

\date{\today}

\begin{abstract}

Feynman's formula for the effective mass of the Fr\"ohlich polaron is rederived from 
the formalism of projected partition functions.  The mass is calculated as inverse of 
the diffusion coefficient of the polaron trajectory in imaginary time.  It is shown 
that correlation between the electron and phonon boundary conditions in imaginary time
is necessary for consistent derivation of the Feynman result.

\end{abstract}


\maketitle

\section{\label{sec:ones}
Introduction
}

Fifty years ago Feynman published his seminal paper \cite{Feynman1955} on the Fr\"ohlich 
polaron, in which he combined path integration and an action variational principle to 
obtain the polaron energy and effective mass.  Subsequently, Feynman's method was 
generalized by many authors \cite{Osaka1959,Abe1971,Luttinger1980,Devreese1992}, for 
comprehensive review see for example Refs.[\onlinecite{Fomin1988,Gerlach1991,Devreese1996}].  
When the exact polaron energy and mass were calculated by Fourier Quantum Monte Carlo
\cite{Alexandrou1990,Alexandrou1992} and Diagrammatic Quantum Monte Carlo 
\cite{Prokof'ev1998,Mishchenko2000} methods, Feynman's polaron energy was found to be 
remarkably accurate deviating from the exact value by less than 0.5\% for all couplings.  
In contrast, Feynman's mass formula \cite{Feynman1955,Schultz1959} {\em over}\/estimated 
the exact mass by as much as 50\% leaving room for possible improvements.  Generalizations 
of the original calculation \cite{Osaka1959,Abe1971,Saitoh1980,Fedyanin1982} produced 
either no or very small, less than 1\%, numerical correction to the mass.  (In some 
instances, the correction was of the wrong sign \cite{Abe1971,Saitoh1980}.)  Thus the  
generalizations did not improve the agreement between the analytical and numerical masses.  
The much larger error in the mass was not regarded as something unexpected.  It was known 
since Feynman's paper that the mass did not satisfy a variational principle, and 
therefore the accuracy of the approximate treatment was expected to be somewhat 
uncontrolled.  

In Feynman's method \cite{Feynman1955,Schultz1963,Feynman1972}, phonon variables are 
integrated out analytically resulting in a self-interacting retarded one-electron action.  
Phonon integration is performed under periodic boundary conditions in imaginary time.  
Periodic boundary conditions are also assumed for the electron trajectory.  Both conditions 
are perfectly appropriate for the energy calculation, since the ground state energy is 
obtained from a full thermodynamic partition function.  However, later, in the mass calculation 
the electron trajectory is broken and its $\tau = 0$ and $\tau = \beta$ ends are displaced 
relative to each other.  The effective mass is inferred from variation of the polaron energy 
with real space separation of the two ends of the trajectory.  In Feynman's approach, the 
retarded polaron action is {\em not} corrected for this displacement.  In other words, the 
action resulted from phonon integration for a periodic electron trajectory is carried over 
to an open electron trajectory.  

In a parallel research on the Holstein polaron 
\cite{KornilovitchPike1997,Kornilovitch1998,Kornilovitch1999} it was realized that the 
boundary conditions for the phonon and electron variables must be synchronized for consistent 
calculation of the polaron mass.  That is, they are either both periodic or both open.  And 
when they are both open they are correlated.  The correlation follows because 
the electron and phonons are coupled in one system and share a common integral of motion, the 
polaron momentum.  Accurate account of the correlation results in a polaron action different 
from the periodic phonon action derived by Feynman.  Thus the question arises of whether this 
difference is significant enough to yield a correction to the Feynman mass formula.  Such a 
correction might account for the deviation from the Monte Carlo mass.  

This question is investigated in the present paper.  It will be shown that the correlation 
between the phonon and electron boundary conditions is essential for consistency of the mass 
calculation.  Without it, a second double integral arising from phonon integration cannot be 
neglected in the mass calculation while it can be neglected in the energy case.  Moreover, it 
spoils the entire mass calculation.  The correlation between the two boundary conditions 
restores internal consistency and leads to the correct final formula.  It will also be shown 
that these observations produce no numerical correction to the Feynman result.

\section{\label{sec:two}
Effective mass from shifted boundary conditions
}

The polaron mass can be calculated as the inverse diffusion coefficient of an open-ended 
polaron path in imaginary time.  This section contains a derivation of this relation.  The 
derivation is general and valid for any non-relativistic composite particle.  The momentum 
$\hbar {\bf K}$ of a translation invariant system is a constant of motion.  In the polaron 
case, this is the sum of an electron momentum and momenta of all excited optical phonons.  
One defines the {\em projected partition function} as a Gibbs sum restricted to states with 
the same {\bf K}: 
\begin{equation}
Z_{\bf K} = \sum_n \langle n \vert e^{-\beta H} \vert n \rangle 
\cdot \delta_{{\bf K} {\bf K}_n} .
\label{eq:one}
\end{equation}
Here $\vert n \rangle$ are eigenstates of the Hamiltonian $H$, $\hbar {\bf K}_n$ is 
the momentum of state $\vert n \rangle$, and $\beta = (k_B T)^{-1}$ is inverse temperature.  
The system is assumed to occupy a finite volume $V$, and the wave vector is quantized to
a discrete set of values ${\bf K}_n$.  To transform $Z_{\bf K}$, introduce real-space 
configurations $\vert Q \rangle$ which are direct products of all the degrees of freedom
in the system.  For the Fr\"ohlich polaron, 
$\vert Q \rangle = \vert {\bf r} \rangle \prod_{{\bf r'}} \vert {\bf P}({\bf r'}) \rangle$,
where ${\bf r}$ is the electron position and ${\bf P}({\bf r'})$ is the polarization at 
point ${\bf r'}$.  The states $Q$ form a complete orthogonal basis, 
$I = \int dQ \vert Q \rangle \langle Q \vert$, and 
$\langle Q_1 \vert Q_2 \rangle = \delta(Q_1 - Q_2)$,
where the unity operator and the delta function are also direct products between different 
degrees of freedom.  Inserting two unity operators in Eq.~(\ref{eq:one}) the partition 
function is rewritten as follows
\begin{equation}
Z_{\bf K} = \int dQ_1 dQ_2 \langle Q_2 \vert e^{- \beta H} \vert Q_1 \rangle 
\cdot W_{\bf K} ,
\label{eq:two}
\end{equation}
\vspace{-0.5cm}
\begin{equation}
W_{\bf K} = \sum_n \langle Q_1 \vert n \rangle \langle n \vert Q_2 \rangle 
\, \delta_{{\bf K} {\bf K}_n} = 
\langle Q_1 \vert {\bf K} \rangle \langle {\bf K} \vert Q_2 \rangle . 
\label{eq:three}
\end{equation}
The meaning of the last expression is that both configurations $Q_1$ and $Q_2$ have to
be projected on the same wave vector ${\bf K}$.  To perform projection consider 
a parallel shift of configuration $Q$ by a three dimensional vector ${\bf R}$.
The resulting state will be denoted as $\vert Q + {\bf R} \rangle$.  (Note that
summation is only symbolic here.)  An arbitrary configuration $Q$ generates a family
of states $\vert {\bf K}_Q \rangle = V^{-1} \int d {\bf R} \: e^{-i {\bf K R}}
\vert Q + {\bf R} \rangle$.  Inversely, $\vert Q + {\bf R} \rangle = 
\sum_{\bf K} e^{i {\bf K R} } \vert {\bf K}_Q \rangle$.
In $W_{\bf K}$ only the respective components $\vert {\bf K}_{Q_1} \rangle$ and 
$\vert {\bf K}_{Q_2} \rangle$ of $\vert Q_1 \rangle$ and $\vert Q_2 \rangle$ survive 
projection on {\bf K}.  As a result, one obtains
\begin{eqnarray}
W_{\bf K} & = & \langle {\bf K}_{Q_1} \vert {\bf K}_{Q_2} \rangle  \nonumber \\
          & = & \frac{1}{V^2} \int d {\bf R}_1 d {\bf R}_2 
\langle Q_1 + {\bf R}_1 \vert Q_2 + {\bf R}_2 \rangle 
e^{i{\bf K}({\bf R}_1 - {\bf R}_2)}                                \nonumber \\
          & = & \frac{1}{V} \int d (\Delta {\bf r}) e^{i{\bf K} \Delta {\bf r}}
\langle Q_1 + \Delta {\bf r} \vert Q_2 \rangle                     \nonumber \\
          & = & \frac{1}{V} \int d (\Delta {\bf r}) e^{i{\bf K} \Delta {\bf r}}
\cdot \delta \left[ (Q_1 + \Delta {\bf r}) - Q_2 \right] ,
\label{eq:four}
\end{eqnarray}
where $\Delta {\bf r} = {\bf R}_1 - {\bf R}_2$.  The resulting delta function ensures 
that a many body configuration $Q_2$ is identical to a configuration $Q_1$ shifted by 
$\Delta {\bf r}$.  Substitution in Eq.~(\ref{eq:two}) and integration over $Q_2$ yield
\begin{equation}
Z_{\bf K} = \frac{1}{V} \int d (\Delta {\bf r}) e^{i{\bf K} \Delta {\bf r}}  
\int dQ \langle Q + \Delta {\bf r} \vert e^{- \beta H} \vert Q \rangle .
\label{eq:five}
\end{equation}
The matrix element under the $dQ$ integral is the density matrix operator taken between 
an arbitrary real-space configuration $Q$ and the {\em same} configuration shifted by 
$\Delta {\bf r}$.  Since $Q$ is a many body state all particles shift in parallel.  In the 
polaron system the electron coordinate {\bf r} and polarization profile ${\bf P}({\bf r'})$ 
shift together.  Equation (\ref{eq:five}) suggests defining the {\em shift partition function}
\begin{equation}
Z_{\Delta {\bf r}} = \int dQ \langle Q + \Delta {\bf r} \vert e^{- \beta H} \vert Q \rangle ,
\label{eq:six}
\end{equation}
which is characterized by the shift vector $\Delta {\bf r}$.  The zero-shift partition 
function, $Z_{\Delta {\bf r} = 0}$, coincides with the usual thermodynamic partition 
function.  Equation (\ref{eq:five}) states that the projected partition function 
and the shift partition function satisfy a Fourier-type relation    
\begin{equation}
Z_{\bf K} = \frac{1}{V} \int d (\Delta {\bf r}) \, e^{i{\bf K} \Delta {\bf r}}  
\cdot Z_{\Delta {\bf r}} .
\label{eq:seven}
\end{equation}
This relation is valid at any temperature $T$.  In the low temperature limit, $Z_{\bf K}$ 
is dominated by the lowest energy eigenvalue with given wave vector, which allows 
derivation of a useful formula for the effective mass.  At small {\bf K}, the energy is
approximated by $E_{\bf K} = E_G + \frac{\hbar^2 {\bf K}^2}{2m^{\ast}}$, and the projected 
partition function $Z_{\bf K} \rightarrow e^{-\beta E_{\bf K}}$ as $T \rightarrow 0$. 
Expanding Eq.~(\ref{eq:seven}) to the second order in ${\bf K}$ one obtains
\begin{eqnarray}
& e^{- \beta E_G} & \!\!\! \left( 1 - \frac{\beta \hbar^2 {\bf K}^2}{2 m^{\ast}} \right) 
\nonumber \\
              & = & \frac{1}{V} \int \! d (\Delta {\bf r}) \! 
\left[ 1 + i ({\bf K} \Delta {\bf r}) - \frac{({\bf K} \Delta {\bf r})^2}{2} \right]   
Z_{\Delta {\bf r}} .
\label{eq:eight}
\end{eqnarray}
On the right, the linear term in {\bf K} vanishes after integration by inversion symmetry: 
$Z_{- \Delta {\bf r}} = Z_{\Delta {\bf r}}$.  The rest is transformed as follows
\begin{equation}
\frac{\beta \hbar^2 {\bf K}^2}{m^{\ast}} = 
\frac{\int d (\Delta {\bf r}) ({\bf K} \Delta {\bf r})^2 Z_{\Delta {\bf r}} }
     {\int d (\Delta {\bf r}) Z_{\Delta {\bf r}} } \equiv
\langle ({\bf K} \Delta {\bf r})^2 \rangle_{\rm shift} .
\label{eq:nine}
\end{equation}
The definition of $Z_{\Delta {\bf r}}$, Eq.~(\ref{eq:six}), implies that the ratio of two 
integrals in the last expression is the mean value of $({\bf K} \Delta {\bf r})^2$ evaluated 
with shifted boundary conditions in imaginary time.  The latter means the initial (at 
imaginary time = 0) and final (at imaginary time = $\beta$) configurations are the same, cf. 
Eq.~(\ref{eq:six}), but they can be shifted relative to each other by a three dimensional 
vector $\Delta {\bf r}$.  This shift vector is arbitrary.  Averaging under shifted boundary 
conditions is understood hereafter as averaging over $\Delta {\bf r}$, with the weight given 
by $Z_{\Delta {\bf r}}$.  Upon expansion of the square in Eq.~(\ref{eq:nine}), the mixed terms 
average to zero by symmetry, $\langle (\Delta r_i) (\Delta r_j) \rangle_{\rm shift} = 0$, while 
the diagonal terms are equal, $\langle (\Delta r_i)^2 \rangle_{\rm shift} = \frac{1}{3} \langle 
(\Delta {\bf r})^2 \rangle_{\rm shift}$.  That results in
\begin{equation}
\frac{1}{m^{\ast}} = \frac{1}{3 \beta \hbar^2}
\frac{\int d (\Delta {\bf r}) (\Delta {\bf r})^2 Z_{\Delta {\bf r}} }
     {\int d (\Delta {\bf r})                    Z_{\Delta {\bf r}} } =
\frac{\langle (\Delta {\bf r})^2 \rangle_{\rm shift} }{3 \beta \hbar^2} .
\label{eq:ten}
\end{equation}
This equation allows an elegant interpretation of the effective mass in terms of 
{\em imaginary time diffusion}.  Since the shift vector is not fixed, the system evolution 
from the initial to the final configuration can be regarded as diffusion during time 
$t = \hbar \beta$.  In normal three-dimensional diffusion, the mean squared displacement
is proportional to the time interval, $\langle (\Delta {\bf r})^2 \rangle = 6 D t$, where
$D$ is the diffusion coefficient.  Thus Eq.~(\ref{eq:ten}) is rewritten as
\begin{equation}
\frac{1}{m^{\ast}} = \frac{2D}{\hbar} .
\label{eq:eleven}
\end{equation}
Note, that Eq.~(\ref{eq:ten}) can also be regarded as a fluctuation-dissipation relation. 
The effective mass characterizes dynamical response, while the mean squared displacement 
is an equilibrium property.   

Equation (\ref{eq:ten}) is especially useful in understanding mass enhancement of composite 
particles such as the polaron.  Interaction with various fields (e.g., phonons) increases 
the statistical weight of trajectories with small $\langle (\Delta {\bf r})^2 \rangle$ 
thereby slowing down the diffusion and increasing the particle's mass.  Thus the mass
enhancement is conveniently visualized as the increased ``stiffness'' of the trajectories.
Equation (\ref{eq:ten}) or its analogues were used in Monte Carlo calculations of the
effective masses of polarons 
\cite{Alexandrou1990,Alexandrou1992,KornilovitchPike1997,Kornilovitch1998,Kornilovitch1999}, 
bipolarons \cite{Macridin}, and defects in superfluid helium \cite{Boninsegni1995}.

\section{\label{sec:three}
Polaron action 
}

The results of the previous section have important implications for the Fr\"{o}hlich polaron.  
As soon as the two ends of the electron path are allowed to shift relative from each other to 
obtain the mass, the polarization profile must shift accordingly.  Thus phonon integration has 
to be performed under more general boundary conditions in imaginary time than periodic.  In 
general, this should modify the polaron action.  The modified polaron action is calculated in 
this section.   

A starting point is the polaron action as formulated by Fr\"ohlich \cite{Froehlich1963} 
and Schultz \cite{Schultz1963}:
\begin{eqnarray}
&S&\!\!\!\![{\bf r}(\tau); {\bf P}({\bf r'},\tau)] = 
- \int^{\beta}_0 d\tau \frac{m {\dot{\bf r}}^2}{2 \hbar^2} + \nonumber \\
 & + & |e| \int^{\beta}_0 d\tau \int d{\bf r'} 
\left( \nabla_{\bf r'} \frac{1}{|{\bf r'} - {\bf r}|} \right) {\bf P}({\bf r'}) - \nonumber \\
 & - & \frac{\mu}{2} \int^{\beta}_0 d\tau \int d{\bf r'} 
\left[ \frac{{\dot{\bf P}}^2({\bf r'})}{\hbar^2} + \Omega^2 {\bf P}^2({\bf r'}) 
\right] ,
\label{eq:twelve}
\end{eqnarray}
\begin{equation}
\mu = \frac{4 \pi}{\Omega^2} \cdot \frac{\varepsilon_0 \: \varepsilon_{\infty}} 
{\varepsilon_0 - \varepsilon_{\infty}} .
\label{eq:thirteen}
\end{equation}
Here $\Omega$ is the optical phonon frequency, $\varepsilon_0$ and $\varepsilon_{\infty}$ 
are the static and high-frequency electric permittivities of the crystal, $|e|$ is the unit 
charge, and $m$ is the band mass of the electron.  ${\bf r}(\tau)$ is the imaginary-time 
electron trajectory.  ${\bf P}({\bf r'},\tau)$ is the imaginary-time polarization trajectory.
A dot above a variable denotes partial derivative with respect to imaginary time $\tau$.  
The action is supplemented by shifted boundary conditions
\begin{eqnarray}
{\bf r}(\beta)           & = & {\bf r}(0) + \Delta {\bf r} ,
\label{eq:fourteen} \\
{\bf P}({\bf r'}, \beta) & = & {\bf P}({\bf r'} - \Delta {\bf r}, 0) ,
\label{eq:fifteen}
\end{eqnarray}
where $\Delta {\bf r}$ is an arbitrary 3-dimensional vector.  Except for the boundary
conditions phonon integration proceeds along the lines outlined by Schultz \cite{Schultz1963}.  
The polarization field is expanded in a Fourier series with real amplitudes $A$ and $B$:
\begin{equation}
{\bf P}({\bf r'},\tau) = \sqrt{\frac{2}{V}} \sum_{(q_x > 0)} 
\frac{{\bf q}}{\vert {\bf q} \vert} \left[ 
A_{\bf q}(\tau) \cos{\bf qr'} + B_{\bf q}(\tau) \sin{\bf qr'} \right] .
\label{eq:sixteen}
\end{equation}
Note that the sum over {\bf q} extends only over half of momentum space, which is 
indicated by `$(q_x > 0)$'.  The Fourier-transformed polaron action (\ref{eq:twelve}) 
and boundary condition (\ref{eq:fifteen}) become 
\begin{eqnarray}
&S&\!\!\![{\bf r}(\tau); A_{\bf q}(\tau), B_{\bf q}(\tau)] = 
- \int^{\beta}_0 d\tau \frac{m {\dot{\bf r}}^2}{2 \hbar^2} \nonumber \\
&+& \sum_{(q_x > 0)} \int^{\beta}_0 d\tau \left\{
- \frac{\mu}{2\hbar^2} \left[ {\dot A}^2_{\bf q} + {\dot B}^2_{\bf q} \right]
- \frac{\mu \Omega^2}{2} \left[ A^2_{\bf q} + B^2_{\bf q} \right] \right. \nonumber \\
&+& \left. \frac{4 \pi |e|}{\vert {\bf q} \vert} \sqrt{\frac{2}{V}} 
\left[ A_{\bf q}(\tau) \sin{\bf qr(\tau)} - B_{\bf q}(\tau) \cos{\bf qr(\tau)} \right]
\!\! \right\} \! ,
\label{eq:seventeen}
\end{eqnarray}
\vspace{-0.3cm}
\begin{eqnarray}
A_{\bf q}(\beta) \cos{{\bf q} \Delta {\bf r}} + 
B_{\bf q}(\beta) \sin{{\bf q} \Delta {\bf r}} & = & A_{\bf q}(0)  
\label{eq:eighteen}  \\
- A_{\bf q}(\beta) \sin{{\bf q} \Delta {\bf r}} + 
  B_{\bf q}(\beta) \cos{{\bf q} \Delta {\bf r}} & = & B_{\bf q}(0)  .
\label{eq:nineteen}
\end{eqnarray}
Since action (\ref{eq:seventeen}) is diagonal in amplitudes $A$ and $B$, path integration 
can be performed for each component independently.  Using the standard methods 
\cite{Feynman1955,Schultz1963,Feynman1972} one obtains in the low-temperature limit 
$e^{\hbar\Omega\beta} \gg 1$:
\begin{widetext}
\begin{eqnarray}
S[{\bf r}(\tau); A_{\bf q}(0), A_{\bf q}(\beta), B_{\bf q}(0), B_{\bf q}(\beta)] = 
- \int^{\beta}_0 d\tau \frac{m {\dot{\bf r}}^2}{2 \hbar^2} + \sum_{(q_x > 0)} S_{\bf q} ,
\makebox[4.0cm]{}
\label{eq:twenty} \\
S_{\bf q} = - \frac{\mu}{2\hbar^2} \, \hbar\Omega \, 
\left[ A^2_{\bf q}(0) + A^2_{\bf q}(\beta) + B^2_{\bf q}(0) + B^2_{\bf q}(\beta) \right] 
+ \frac{\hbar^2}{2\mu} \frac{(4 \pi |e|)^2}{q^2} \frac{2}{V}
\int^{\beta}_0 \!\!\! \int^{\beta}_0 d\tau' d\tau''  G(\tau',\tau'') 
\cos{{\bf q}[{\bf r}(\tau')-{\bf r}(\tau'')]} +  \nonumber \\
+ \frac{(4 \pi |e|)}{q} \sqrt{\frac{2}{V}} \!\! \int^{\beta}_0 \!\!\!\! d\tau 
\left\{
e^{-\hbar\Omega\tau} \left[ A_{\bf q}(0) \sin{{\bf q}{\bf r}(\tau)} 
                          - B_{\bf q}(0) \cos{{\bf q}{\bf r}(\tau)} \right] +
e^{-\hbar\Omega(\beta-\tau)} \left[ A_{\bf q}(\beta) \sin{{\bf q}{\bf r}(\tau)} 
                                  - B_{\bf q}(\beta) \cos{{\bf q}{\bf r}(\tau)} \right]
\right\} , \\
G(\tau',\tau'') = \frac{1}{\hbar\Omega \sinh{\hbar\Omega\beta}} \cdot \left\{
\begin{array}{ll}
\sinh{\hbar\Omega\tau'} \cdot \sinh{\hbar\Omega(\beta-\tau'')}; & \tau' < \tau'' \\
\sinh{\hbar\Omega(\beta-\tau')} \cdot \sinh{\hbar\Omega\tau''}; & \tau' > \tau''
\end{array} . \right.  \makebox[4.0cm]{} 
\label{eq:twentytwo}
\end{eqnarray}
The action is still a functional of two end polarizations, at $\tau = 0$ and $\tau = \beta$.  
However, those are related by the conditions (\ref{eq:eighteen})-(\ref{eq:nineteen}).  
Final integration over the end variables leads, after straightforward algebra, to
\begin{equation}
S_{\Delta {\bf r}}[{\bf r}(\tau)] = 
- \int^{\beta}_0 d\tau \frac{m {\dot{\bf r}}^2}{2 \hbar^2} + \frac{\alpha}{2\sqrt{2}} 
\left( \frac{\hbar^5 \Omega^3}{m} \right)^{\!\!\frac{1}{2}}
\!\! \int^{\beta}_0 \!\!\! \int^{\beta}_0 \!\! d\tau' d\tau'' \left\{ 
\frac{e^{-\hbar\Omega |\tau'-\tau''|}}{\vert {\bf r}(\tau') - {\bf r}(\tau'') \vert} +
\frac{e^{-\hbar\Omega (\beta - |\tau'-\tau''|)}}
{\vert [{\bf r}(\tau') - {\bf r}(\tau'')] \: {\rm sgn}(\tau'-\tau'') - \Delta {\bf r} \vert}
\right\} ,
\label{eq:twentythree}
\end{equation}
\begin{equation}
\alpha = \frac{|e|^2}{2 \hbar \Omega} 
\left( \frac{1}{\varepsilon_{\infty}} - \frac{1}{\varepsilon_0} \right)
\sqrt{\frac{2 m \Omega}{\hbar}} ,
\label{eq:twentyfour}
\end{equation}
where $\alpha$ is the Fr\"ohlich coupling constant.  $\Delta {\bf r}$ in the denominator
of the last term in Eq.~(\ref{eq:twentythree}) is a direct consequence of the shifted 
boundary conditions for the polarization field.  It is convenient to perform a linear 
transformation ${\bf r}(\tau) = {\bf r}'(\tau) + \frac{\tau}{\beta} \Delta {\bf r}$, which
transforms path integration to periodic boundary conditions.  The resulting action is
\begin{eqnarray}
{\bar S}_{\Delta {\bf r}}[{\bf r}(\tau)] = 
- \frac{m}{2\hbar^2} \frac{(\Delta {\bf r})^2}{\beta}
- \int^{\beta}_0 d\tau \frac{m {\dot{\bf r}}^2}{2 \hbar^2} +
\nonumber \\
+ \frac{\alpha}{2\sqrt{2}} \left( \frac{\hbar^5 \Omega^3}{m} \right)^{\!\!\frac{1}{2}}
\int^{\beta}_0 \hspace{-2.0mm} \int^{\beta}_0 \hspace{-2.0mm} d\tau' d\tau'' 
\left\{ \frac{e^{-\hbar\Omega |\tau'-\tau''|}}
{\vert {\bf r}(\tau') - {\bf r}(\tau'') + \frac{\tau'-\tau''}{\beta} \Delta {\bf r} \vert} 
+  \frac{e^{-\hbar\Omega (\beta - |\tau'-\tau''|)}}
{\vert [{\bf r}(\tau') - {\bf r}(\tau'')] \: {\rm sgn}(\tau'-\tau'') - 
\frac{\beta - \vert \tau'-\tau'' \vert}{\beta} \Delta {\bf r} \vert} \right\} .  
\label{eq:twentyfive}
\end{eqnarray}
\end{widetext}
The bar over $S$ indicates that the action is a functional of a path periodic in imaginary 
time.  The full shift partition function (\ref{eq:six}) is given by 
\begin{equation}
Z_{\Delta {\bf r}} = Z_{\rm ph} \cdot \int^{({\bf r}, \, \beta)}_{({\bf r}, \, 0)}
{\cal D} {\bf r} \cdot e^{{\bar S}_{\Delta {\bf r}}[ \: {\bf r}(\tau)] } ,
\label{eq:twentysix}
\end{equation}
where ${\cal D}{\bf r}$ is path integration over electron coordinates.  $Z_{\rm ph}$ is the 
partition function of a free polarization field.  This is a multiplicative constant that 
cancels out in the mass calculation, cf. Eq.~(\ref{eq:ten}).   

The polaron mass originates from explicit dependence of action (\ref{eq:twentyfive}) on 
the shift vector $\Delta {\bf r}$.  The first term corresponds to the bare electron mass.  
Phonon-induced mass enhancement comes from the double integral.  The two integrands have 
similar functional dependence on ${\bf r}(\tau)$, but make different contributions to the 
action.  The first integrand exponentially decays away from the diagonal $\tau' = \tau''$.  
Since the odd powers of $\Delta {\bf r}$ vanish in path integration, the first fraction's 
contribution is ${\cal O}(\beta) + {\cal O}((\Delta {\bf r})^2 \beta^{-1}) + 
{\cal O}((\Delta {\bf r})^4 \beta^{-3}) + \ldots$  The first term of this expansion adds 
to the polaron energy, while the second one adds to the mass.  

Consider the second fraction in the double integral (\ref{eq:twentyfive}), which is the main 
focus of the present study.  The exponential numerator limits integration to finite intervals 
around the points $(0,\beta)$ and $(\beta,0)$.  In the denominator, the 
combination $(\beta - |\tau'-\tau''|) = {\cal O}(1)$.  Therefore the second fraction's 
contribution to the action is ${\cal O}(1) + 
{\cal O}((\Delta {\bf r})^2 \beta^{-2}) + {\cal O}((\Delta {\bf r})^4 \beta^{-4}) + \ldots$
In the $\beta \rightarrow \infty$ limit, each term in this expansion is small in comparison
with the corresponding term from the previous series with the same power of $\Delta {\bf r}$.
Thus the {\em entire} second fraction can be omitted in favor of the first one.  

Here comes the critical observation.  Such a nice term-by-term domination of the first 
fraction over the second one takes place {\em only} as a result of the shifted boundary 
conditions for the polarization.  Indeed, without the latter the combination in the 
denominator of the second fraction in Eq.~(\ref{eq:twentyfive}) would have been 
$\frac{|\tau'-\tau''|}{\beta} \Delta {\bf r} = {\cal O}(\Delta {\bf r})$ instead of 
$\frac{(\beta - |\tau'-\tau''|)}{\beta} \Delta {\bf r} = {\cal O}((\Delta {\bf r}) \beta^{-1})$.  
As a result, the second expansion would have been ${\cal O}(1) + {\cal O}((\Delta {\bf r})^2) 
+ {\cal O}((\Delta {\bf r})^4) + \ldots$  The first term is still small compared to the
corresponding term from the first fraction, which leads to the correct polaron energy.
However, the $(\Delta {\bf r})^2$ term now dominates its counterpart from the first fraction,
which totally confuses calculation of the effective mass.  Thus the neglect of the shifted
boundary conditions in phonon integration results in a serious internal inconsistency in the 
mass calculation.  Apparently, Feynman avoided this difficulty by omitting the second fraction 
in Eq.~(\ref{eq:twentythree}) from the outset.  Had he retained the full form of the phonon 
propagator, including the second part $e^{-\hbar\Omega(\beta - |\tau'-\tau''|)}$, he would 
have faced the problem outlined here.    

The analysis presented in this section enables to safely neglect the second fraction in  
Eq.~(\ref{eq:twentyfive}).  However, it will still be included in the forthcoming mass 
calculation in order to illustrate further the above argument.

\section{\label{sec:four}
Polaron mass
}

Polaron action (\ref{eq:twentyfive}) is real.  Therefore the shifted partition function
satisfies the Jensen-Feynman inequality 
\begin{eqnarray}
\int^{{\bf r}_0}_{{\bf r}_0}
{\cal D} {\bf r} \cdot e^{{\bar S}_{\Delta {\bf r}}[{\bf r}(\tau)] } \geq
e^{\langle {\bar S}_{\Delta {\bf r}}[{\bf r}(\tau)] - 
         {\bar S}^0_{\Delta {\bf r}}[{\bf r}(\tau)] 
\rangle_0} \cdot \nonumber \\
\cdot \int^{{\bf r}_0}_{{\bf r}_0}
{\cal D} {\bf r} \cdot e^{{\bar S}^0_{\Delta {\bf r}}[{\bf r}(\tau)] } ,
\label{eq:twentyseven}
\end{eqnarray}
where $\langle ... \rangle_0$ denotes averaging with the trial action ${\bar S}_0$.  Feynman's 
trial model consists of two particles with masses $m$ and $M$, which are elastically coupled  
with a spring constant $\kappa$.  Note that if the first particle's mass is different from the 
electron mass $m$, the difference $\langle {\bar S} - {\bar S}_0 \rangle_0$ diverges.  The 
second particle's mass $M$ and $\kappa$ are variational parameters.  It is customary to replace 
them with two parameters $w = \hbar \sqrt{\frac{\kappa}{M}}$ and $v = w \sqrt{1 + \frac{M}{m}}$
which both have units of energy.  Calculation of the right-hand side of Eq.~(\ref{eq:twentyseven}) 
is tedious, but proceeds along essentially the same lines as the original Feynman calculation.  
Therefore the intermediate steps are not presented here.  An important note concerns the 
$\Delta {\bf r}$ dependence of the trial action.  It is fully represented by the term 
$- \frac{(m+M)}{2\hbar^2} \frac{(\Delta {\bf r})^2}{\beta}$, reflecting the fact that the 
total mass of the trial model is $m + M$.  As a result of the calculation, inequality 
(\ref{eq:twentyseven}) takes the form
\begin{widetext}
\begin{equation}
Z_{\Delta {\bf r}} \geq Z_{\rm ph} \cdot 
\left( \frac{m}{2 \pi \hbar^2 \beta} \right)^{\frac{3}{2}} \cdot
\exp \left\{ - \beta E_F + {\cal O}(1) - \frac{m_F}{2\hbar^2} \frac{(\Delta {\bf r})^2}{\beta} + 
{\cal O}((\Delta {\bf r})^4 \beta^{-3}) + \ldots \right\}  ,
\label{eq:twentyeight}
\end{equation}
\begin{equation}
E_F = \frac{3}{4} \frac{(v-w)^2}{v} - \frac{\alpha}{2\sqrt{\pi}} (\hbar \Omega)^{\frac{3}{2}}
\frac{1}{\beta} \int^{\beta}_{0} \hspace{-0.2cm} \int^{\beta}_{0} d\tau' d\tau'' 
\frac{e^{- \hbar \Omega \vert \tau'-\tau'' \vert} + 
      e^{- \hbar \Omega (\beta - \vert \tau'-\tau'' \vert)} }
{[\Phi(\tau',\tau'')]^{\frac{1}{2}}} ,
\label{eq:twentynine}
\end{equation}  
\begin{equation}
m_F = m \left\{ 1 + \frac{\alpha}{6\sqrt{\pi}} (\hbar \Omega)^{\frac{3}{2}}
\frac{1}{\beta} \int^{\beta}_{0} \hspace{-0.2cm} \int^{\beta}_{0} d\tau' d\tau'' 
\frac{\vert \tau'-\tau'' \vert^2 e^{- \hbar \Omega \vert \tau'-\tau'' \vert} + 
(\beta - \vert \tau'-\tau'' \vert)^2 e^{- \hbar \Omega (\beta - \vert \tau'-\tau'' \vert)} }
{[\Phi(\tau',\tau'')]^{\frac{3}{2}}}
\right\} ,
\label{eq:thirty}
\end{equation}  
\begin{eqnarray}
\Phi(\tau',\tau'') = \frac{1}{1 + \frac{M}{m}} 
\left\{ \vert \tau'-\tau'' \vert - \frac{( \tau'-\tau'' )^2}{\beta} \right\} 
& + & \frac{1}{v \left( 1 + \frac{m}{M} \right)} 
\left\{ 1 - \frac{1}{2} \: e^{-2v\tau'}  - \frac{1}{2} \: e^{-2v(\beta-\tau')} 
\right.
\nonumber \\
- \frac{1}{2} \: e^{-2v\tau''} - \frac{1}{2} \: e^{-2v(\beta-\tau'')} 
& + & \left. e^{-v(\tau'+\tau'')} + e^{-v[2\beta-(\tau'+\tau'')]} 
- e^{-v |\tau'-\tau''|} \right\} .
\label{eq:thirtyone}
\end{eqnarray}  
\end{widetext}
$E_F$ and $m_F$ stand for the Feynman energy and Feynman mass, respectively.  These
definitions are understood as functional dependencies only, the optimal values of 
parameters $v$ and $w$ are yet to be determined.  Both $E_F$ and $m_F$ contain terms 
that originate from the last fraction in the action (\ref{eq:twentyfive}).  The critical 
question is about the order of their contribution in the low temperature limit 
$\beta \rightarrow \infty$.  In Eq.~(\ref{eq:twentynine}) the two double integrals are 
${\cal O}(\beta)$ and ${\cal O}(1)$, respectively.  The second one can therefore be 
omitted.  The same is true about the two double integrals in Eq.~(\ref{eq:thirty}).    
In the second one, the integration region is limited to where the combination 
$(\beta - \vert \tau' - \tau'' \vert) \leq (\hbar \Omega)^{-1} = {\cal O}(1)$.  Since 
the pre-exponential factor is exactly the same combination squared, it is ${\cal O}(1)$ 
throughout the essential integration region.  As a result, the entire second double 
integral is ${\cal O}(1)$.  Therefore it can be omitted in favor of the first integral,
which is ${\cal O}(\beta)$.  Now recall that the pre-exponential factor 
$(\beta - \vert \tau' - \tau'' \vert)^2$ derives from the shifted boundary conditions
of the phonon integration.  Without the latter, the pre-exponential factor would have
been $\vert \tau' - \tau'' \vert^2$, that is the same as in the first integral.  
Therefore, it would have been ${\cal O}(\beta^2)$ in the essential integration region.
That would have resulted in the second integral being ${\cal O}(\beta^2)$ and its
contribution to the mass ${\cal O}(\beta)$, which, of course, makes no sense.  Thus 
the shifted boundary conditions in phonon integration are essential for the correct
form of the pre-exponential factor and, in the end, for the consistency of the entire
mass calculation.  Once the second integrals in Eqs.~(\ref{eq:twentynine}) and
(\ref{eq:thirty}) are safely omitted, the first ones can be transformed according to
the relation  
\begin{equation}
\int^{\beta}_{0} \hspace{-0.2cm} \int^{\beta}_{0} d\tau' d\tau''
g(\vert \tau' - \tau'' \vert) \approx 2 \beta \int^{\beta}_{0} d\tau g(\tau) , 
\label{eq:thirtytwo}
\end{equation}  
which is valid for any function $g$ in the limit $\beta \rightarrow \infty$.  
The Feynman energy and mass assume their final forms: 
\begin{equation}
E_F = \frac{3}{4} \frac{(v-w)^2}{v} - \frac{\alpha}{\sqrt{\pi}} (\hbar \Omega)^{\frac{3}{2}}
\int^{\infty}_{0} d\tau \frac{e^{- \hbar \Omega \tau}}{[F(\tau)]^{\frac{1}{2}}} ,
\label{eq:thirtythree}
\end{equation}  
\begin{equation}
m_F = m \left\{ 1 + \frac{\alpha}{3\sqrt{\pi}} (\hbar \Omega)^{\frac{3}{2}}
\int^{\infty}_{0} d\tau \frac{\tau^2 e^{- \hbar \Omega \tau}}{[F(\tau)]^{\frac{3}{2}}} 
\right\} ,
\label{eq:thirtyfour}
\end{equation}  
\begin{equation}
F(\tau) = \frac{w^2}{v^2} \tau + \frac{v^2-w^2}{v^3} \left( 1 - e^{-v\tau} \right) .
\label{eq:thirtyfive}
\end{equation}  

The last step is choosing optimal values of $v$ and $w$.  The standard approach has always 
been to minimize $E_F$ first and then substitute the obtained values into the expression for 
$m_F$.  However, the variational theorem (\ref{eq:twentyeight}) is valid for any $\Delta {\bf r}$ 
and not for just the zero shift vector.  The question therefore is whether the other terms in 
the polaron action change the optimal values of the variational parameters.  Let $v_0$ and 
$w_0$ minimize function $E_F(v,w)$ defined by Eq.~(\ref{eq:thirtythree}).  As such, they 
satisfy the equations $\partial E_F/\partial v = 0$ and $\partial E_F/\partial w = 0$ and a 
necessary concavity condition.  With other terms included, one has to minimize 
$E_F + {\rm const}/\beta + (m_F/2\hbar^2)(\Delta {\bf r}/\beta)^2 + \ldots$  Note that the 
second and third terms are of the same order ${\cal O}(\beta^{-1})$.  Minimization yields 
$v = v_0 + \delta v$, and $w = w_0 + \delta w$, where 
$\delta v, \delta w = {\cal O}(\beta^{-1})$.  Some parts of the corrections $\delta v$ and 
$\delta w$ depend explicitly on $\Delta {\bf r}$.  However, the {\em minimum energy itself} 
receives a correction that is only quadratic in $\delta v$ and $\delta w$.  The leading 
$\Delta {\bf r}$ correction to the minimal polaron action is 
${\cal O}((\Delta {\bf r}/\beta)^2)$, which does not affect the effective mass term 
$-(m_F/2\hbar^2)(\Delta {\bf r})^2/\beta$ in the $\beta \rightarrow \infty$ limit.  

To conclude, the polaron mass is still determined by the original Feynman procedure:
minimize the energy (\ref{eq:thirtythree}) first, and then use the optimal parameters
to calculate the mass from Eq.~(\ref{eq:thirtyfour}).

\section{\label{sec:six}
Summary
}

In this paper, calculation of the Fr\"{o}hlich polaron effective mass has been analyzed for 
robustness to boundary conditions in imaginary time.  It has been shown that a consistent mass 
calculation must involve a correlation between the boundary conditions of the electron and 
phonon coordinates.  The $\tau = \beta$ end points of all paths have to be shifted by the same 
vector $\Delta {\bf r}$ relative to their $\tau = 0$ end points.  Then the effective mass is 
found as inverse diffusion coefficient of the many body path, cf. Eq.~(\ref{eq:eleven}), where 
$\Delta {\bf r}$ is used as a diffusion distance and $\beta \rightarrow \infty$ as a diffusion 
time.  This conclusion is not limited to the polaron system but is valid for any composite 
non-relativistic quantum particle.  
   
It has also been shown that the correlation between the electron and phonon boundary conditions 
is critical for the consistency of the polaron mass calculation.  It results in the correct 
form of a prefactor in the intermediate expression for mass, Eq.~(\ref{eq:thirty}), which 
allows dropping the second double integral altogether.  In the original Feynman calculation, 
the second double integral was omitted from the beginning, i.e., during the energy calculation, 
and therefore did not to cause any problems in the mass calculation.  Such an approach was 
internally inconsistent.  It has been shown here that the consistency is restored via correct 
treatment of the shifted boundary conditions.    

Finally, it has been shown that all the above considerations do not change the numerical 
values of the polaron effective mass obtained in \cite{Feynman1955,Schultz1959}.  The optimal 
values of the variational parameters are still determined by minimization of the ground state
energy (\ref{eq:thirtythree}) alone.  The effective mass then follows from expression
(\ref{eq:thirtyfour}) evaluated at the optimal values of $v$ and $w$.

\end{document}